\documentclass[12pt]{article}
\usepackage{color}
\usepackage{amssymb}
\usepackage{amsmath}
\usepackage{pifont}
\usepackage
[dvips,
colorlinks=true,
linkcolor=webgreen,%defined below
filecolor=webbrown,%defined below
citecolor=webgreen,%defined below
bookmarksopen=false,
]{hyperref}
\definecolor{webgreen}{rgb}{0,.5,0}
\definecolor{webbrown}{rgb}{.6,0,0}
\usepackage[T1]{fontenc}
\usepackage[cp1250]{inputenc}
\usepackage{array}
\usepackage[dvips]{graphicx}
\usepackage{amssymb}
\date{}
\definecolor{arcolor}{cmyk}{0.05,0.95,0.9,0.1}
\title{Quantum diffusion of prices and profits}
\author{Edward W. Piotrowski\\ Institute of Theoretical Physics,
University of Bia\l ystok,\\ Lipowa 41, Pl 15424 Bia\l ystok,
Poland\\ e-mail: \href{mailto:ep@alpha.uwb.edu.pl}{ep@alpha.uwb.edu.pl}\\
 Jan S\l adkowski\\ Institute of Physics, University of Silesia, \\ Uniwersytecka
4, Pl 40007 Katowice, Poland \\ e-mail:
\href{mailto:sladk@us.edu.pl}{sladk@us.edu.pl} }
\begin{document}
\baselineskip9mm
\maketitle
\def\Z{{\bf Z\!\!Z}}
\def\R{{\bf I\!R}}
\def\N{{\bf I\!N}}
\begin{abstract}
\baselineskip7mm We discuss the time evolution of quotations of
stocks and commodities and show that  corrections to the orthodox
Bachelier model inspired by quantum mechanical time evolution of
particles may be important. Our analysis shows that traders
tactics can interfere as waves do  and trader's strategies can be
reproduced from the corresponding Wigner functions. The proposed
interpretation of the chaotic movement of market prices imply that
the Bachelier behaviour follows from short-time interference of
tactics adopted (paths followed) by the rest of the world
considered as a single trader and the Ornstein-Uhlenbeck
corrections to the Bachelier model should qualitatively matter
only for large time scales. The famous smithonian invisible hand
is interpreted as a short-time tactics of whole the market
considered as a single opponent. We also propose a solution to the
currency preference paradox.
\end{abstract}

% PACS numbers: 02.50.-r, 02.50.Le, 03.67.-a, 03.65.Bz
 \vspace{5mm}

\section{Introduction}

We have  formulated a new approach to quantum game theory
\cite{1}-\cite{3} that is suitable for description of market
transactions in term of supply and demand curves
\cite{4}-\cite{8}. In this approach quantum strategies are vectors
(called states) in some Hilbert space and can be interpreted as
superpositions of trading decisions. Tactics or moves are
performed by unitary transformations on vectors in the Hilbert
space (states). The idea behind using quantum games is to explore
the possibility of forming linear combination of amplitudes that
are complex Hilbert space vectors  (interference, entanglement
\cite{3}) whose squared absolute values give probabilities of
players actions. It is generally assumed that a physical
observable (e.g\mbox{.} energy, position), defined by the
prescription for its measurement, is represented by a linear
Hermitian operator. Any measurement of an observable produces an
eigenvalue of the operator representing the observable with some
probability. This probability is given by the squared modulus of
the coordinate corresponding to this eigenvalue in the spectral
decomposition of the state vector describing the system. This is
often an advantage over classical probabilistic description where
one always deals directly with probabilities. The formalism has
potential applications outside physical laboratories \cite{4}.
Strategies and not the apparatus or installation for actual
playing  are at the very core of the approach. Spontaneous or
institutionalized market transactions are described in terms of
projective operation acting on Hilbert spaces of strategies of the
traders. Quantum entanglement is necessary (non-trivial linear
combinations of vectors-strategies have to be formed) to strike
the balance of trade. This approach predicts the property of
undividity of attention of traders (no cloning theorem) and
unifies  the English auction with the Vickrey's one attenuating
the motivation properties of the later \cite{5}. Quantum
strategies create unique opportunities for making profits during
intervals shorter than the characteristic thresholds for an
effective market (Brown motion) \cite{5}. On such market prices
correspond to Rayleigh particles approaching equilibrium state.
Although the effective market hypothesis assumes immediate price
reaction to new information concerning the market the information
flow rate is limited by physical laws such us the constancy of the
speed of light. Entanglement of states allows to apply quantum
protocols of super-dense coding \cite{6} and get ahead of
"classical trader". Besides, quantum version of the famous Zeno
effect \cite{4} controls the process of reaching the equilibrium
state by the market.  Quantum arbitrage based on such phenomena
seems to be feasible. Interception of profitable quantum
strategies is forbidden by the impossibility of cloning of quantum
states.

There are apparent analogies with quantum thermodynamics that
allow to interpret market equilibrium as a state with vanishing
financial risk flow. Euphoria, panic or herd instinct often cause
violent changes of market prices. Such phenomena can be described
by non-commutative quantum mechanics. A simple tactics that
maximize the trader's profit on an effective market follows from
the model: {\it accept profits equal or greater than the one
you have formerly achieved on average}\/ \cite{7}. \\
The player strategy $|\psi\rangle$\footnote{We use the standard
Dirac notation. The symbol $|\ \rangle$ with  a letter $\psi$ in
it denoting a vector parameterized by $\psi$ is called a {\it
ket}; the symbol $\langle\ |\negthinspace$ with a letter in it is
called a {\it bra}. Actually a {\it bra} is a dual vector to the
corresponding {\it ket}. Therefore scalar products of vectors take
the form $\langle \phi |\psi\rangle\negthinspace$ ({\it bracket})
and the expectation value of an operator $A$ in the state
$|\psi\rangle\negthinspace$ is given by $\langle \psi
|A\psi\rangle\negthinspace$. } belongs to some Hilbert space and
have two important representations $\langle
q|\psi\rangle\negthinspace$ (demand representation) and $\langle
p|\psi\rangle\negthinspace$ (supply representation) where $q$ and
$p$ are logarithms of prices at which the player is buying or
selling, respectively \cite{4,8}. After consideration of the
following facts:
\begin{itemize}
\item error theory: second moments of a random
variable describe errors
\item M. Markowitz's portfolio theory
\item L. Bachelier's theory of options:  the random variable $q^{2} + p^{2}$ measures joint risk
for a stock buying-selling transaction ( and Merton \& Scholes
works that gave them Nobel Prize in 1997)
\end{itemize}

we have defined canonically conjugate Hermitian operators
(observables) of demand $\mathcal{Q}_k$ and supply $\mathcal{P}_k$
corresponding to the variables $q$ and $p$ characterizing strategy
of  the k-th player. This led us to the definition of the
observable that we call {\it the risk inclination operator}:
$$
H(\mathcal{P}_k,\mathcal{Q}_k):=\frac{(\mathcal{P}_k-p_{k0})^2}{2\,m}+
                     \frac{m\,\omega^2(\mathcal{Q}_k-q_{k0})^2}{2}\,,
\label{hamiltonian} $$ \noindent where
$p_{k0}\negthinspace:=\negthinspace\frac{
\phantom{}_k\negthinspace\langle\psi|\mathcal{P}_k|\psi\rangle_k }
{\phantom{}_k\negthinspace\langle\psi|\psi\rangle_k}\,$,
$q_{k0}\negthinspace:=\negthinspace\frac{
\phantom{}_k\negthinspace\langle\psi|\mathcal{Q}_k|\psi\rangle_k }
{\phantom{}_k\negthinspace\langle\psi|\psi\rangle_k}\,$,
$\omega\negthinspace:=\negthinspace\frac{2\pi}{\theta}\,$.  $
\theta$ denotes the characteristic time of transaction \cite{7,8}
which is, roughly speaking, an average time spread between two
opposite moves of a player (e.~g.~buying and selling the same
commodity). The parameter $m\negthinspace>\negthinspace0$ measures
the risk asymmetry between buying and selling positions. Analogies
with quantum harmonic oscillator allow for the following
characterization of quantum market games. One can introduce the
constant $h_E$ that describes the minimal inclination of the
player to risk, $
[\mathcal{P}_k,\mathcal{Q}_k]=\frac{i}{2\pi}h_E$. As the lowest
eigenvalue of the positive definite operator $H$ is
$\frac{1}{2}\frac{h_E}{2\pi} \omega$, $h_E$ is equal to the
product of the lowest eigenvalue of
$H(\mathcal{P}_k,\mathcal{Q}_k) $ and $2\theta$. $2\theta $ is in
fact the minimal interval during which it makes sense to measure
the profit.
%\section{Quantum tomography}
Let us consider a simple market with a single commodity
$\mathfrak{G}$. A consumer (trader) who buys this commodity
measures his/her profit in terms of the variable
$\mathfrak{w}\negthinspace=\negthinspace-\mathfrak{q}$.
 The producer who provides the consumer with the
commodity uses
$\mathfrak{w}\negthinspace=\negthinspace-\mathfrak{p}$ to this
end. Analogously, an auctioneer uses the variable
$\mathfrak{w}\negthinspace=\negthinspace\mathfrak{q}$ (we neglect
the additive or multiplicative constant  brokerage) and a
middleman who reduces the store and sells twice as much as he buys
would use the variable
$\mathfrak{w}\negthinspace=\negthinspace-2\hspace{.1em}\mathfrak{p}-\mathfrak{q}$.
Various subjects active on the market may  manifest different
levels of activity. Therefore it is useful to define a standard
for the "canonical" variables $\mathfrak{p}$ and $\mathfrak{q}$ so
that the risk variable \cite{8} takes the simple form
$\tfrac{\mathfrak{p}^2}{2}\negthinspace+\negthinspace\tfrac{\mathfrak{q}^2}{2}$
and the variable $\mathfrak{w}$ measuring the profit of a concrete
market subject dealing in the commodity $\mathfrak{G}$ is given by
\begin{equation}
u\,\mathfrak{q}+v\,\mathfrak{p}+\mathfrak{w}(u,v)=0\,,
\label{rowprosrad}
\end{equation}
where the parameters $u$ and $v$ describe the activity. The dealer
can modify his/her strategy $|\psi\rangle$ to maximize the profit
but this should be done within the specification characterized by
$u$ and $v$. For example, let us consider a fundholder who
restricts himself to purchasing realties. From his point of view,
there is no need nor opportunity of modifying the supply
representation of his strategy because this would not increase the
financial gain from the purchases. One can easily show by
recalling the explicit form of the probability amplitude
$|\psi\rangle\negthinspace\in\negthinspace\mathcal{L}^2$ that the
triple $(u,v,|\psi\rangle)$ describes properties of the profit
random variable $\mathfrak{w}$ gained from trade in the commodity
$\mathfrak{G}$. We will use the Wigner function $W(p,q)$ defined
on the phase space $(p,q)$
\begin{eqnarray*}
W(p,q)&:=& h^{-1}_E\int_{-\infty}^{\infty}e^{i\hslash_E^{-1}p x}
\;\frac{\langle
q+\frac{x}{2}|\psi\rangle\langle\psi|q-\frac{x}{2}\rangle}
{\langle\psi|\psi\rangle}\; dx\\
&=& h^{-2}_E\int_{-\infty}^{\infty}e^{i\hslash_E^{-1}q x}\;
\frac{\langle
p+\frac{x}{2}|\psi\rangle\langle\psi|p-\frac{x}{2}\rangle}
{\langle\psi|\psi\rangle}\; dx,
\end{eqnarray*}
to measure the (pseudo-)probabilities of the players behaviour
implied by his/her strategy $|\psi\rangle$ (the positive constant
$h_E=2\pi\hslash_E$ is the dimensionless economical counterpart of
the Planck constant discussed in the previous section \cite{4,8}).
Therefore if we fix values of the parameters $u$ and $v$ then  the
probability distribution of the random variable $\mathfrak{w}$ is
given by a marginal distribution $W_{u,v}(w)dw$ that is equal to
the Wigner function $W(p,q)$ integrated over the line
$u\,p+v\,q+w=0$:
\begin{equation}
W_{u,v}(w):=\iint\displaylimits_{\mathbb{R}^2}W(p,q)\,\delta(u\,
q\negthinspace+\negthinspace v\,p\,\negthinspace+\negthinspace
w,0) \,dpdq\, , \label{defiont}
\end{equation}
where the Dirac delta function is used to force the constraint (
$\delta(u\, q\negthinspace+\negthinspace
v\,p\,\negthinspace+\negthinspace w,0)$).  The above integral
transform $(W\negthinspace:\negthinspace\mathbb{R}^2
\negthinspace\rightarrow\negthinspace \mathbb{R})\longrightarrow
(W\negthinspace:\negthinspace\mathbb{P}^2\negthinspace\rightarrow\negthinspace\mathbb{R})$
is known as the Radon transform \cite{9}. Let us note that the
function $W_{u,v}(w)$ is homogeneous of the order -1, that is
\begin{equation*}
W_{\lambda u,\lambda v}(\lambda w)=|\lambda|^{-1}W_{u,v}(w)\,.
\end{equation*}
Some special examples of the (pseudo-) measure $W_{u,v}(w)dw$
where previously discussed in \cite{4,8,10}. The squared absolute
value of a pure strategy in the supply representation is equal to
$W_{0,1}(p)$ ($|\langle p|\psi\rangle|^2=W_{0,1}(p)$) and in the
demand representation the relation reads $|\langle
q|\psi\rangle|^2=W_{1,0}(q)$. It is positive definite in these
cases for all values of $u$ and $v$. If we express the variables
$u$ and $v$ in the units $\hslash_E^{-\frac{1}{2}}$ then the
definitions of $W(p,q)$ and $W_{u,v}$ lead to the following
relation between $W_{u,v}(w)$ i $\langle p|\psi\rangle$ or
$\langle q|\psi\rangle$  for both representations\footnote{One
must remember that switching roles of $p$ and $q$ must be
accompanied by switching $u$ with $v$} \cite{11}:
\begin{equation} W_{u,v}(w)\frac{1}{2\pi|v|}\Bigl|\int_{-\infty}^\infty\negthinspace\negthinspace
\text{e}^{\frac{\text{i}}{2v}(up^2+2pw)}\langle p|\psi\rangle\,
dp\,\Bigr|^2. \label{radonpsi}
\end{equation}
The integral representation of the Dirac delta function
 \begin{equation}
\delta(uq\negthinspace+\negthinspace vp\negthinspace+\negthinspace
w,0)=\frac{1}{2\pi}\int^\infty_{-\infty}\text{e}^{
\text{i}k(uq+vp+w)}dk \label{hraddel}
\end{equation}
helps with finding the reverse transformation to
$(\ref{defiont})$. The results is:
\begin{equation}
W(p,q)=\frac{1}{4\pi^2}\iiint\displaylimits_{\mathbb{R}^3}\cos(uq\negthinspace+\negthinspace
vp\negthinspace+\negthinspace w)\, W_{u,v}(w)\,dudvdw\,.
\label{odwrrado}
\end{equation}
Traders  using the same strategy (or  single traders that can
adapt their moves to variable market situations) can form sort of
"tomographic pictures" of their strategies by measuring profits
from trading in the commodity $\mathfrak{G}$. These pictures would
be influenced by various circumstances and characterized by values
of $u$ and $v$. These data can be used for reconstruction of the
respective strategies expressed in terms the Wigner functions
$W(p,q)$ according to the formula $(\ref{odwrrado})$.
\subsection{Example: marginal distribution of an adiabatic
strategy} Let us consider the Wigner function of the $n$-th
excited\footnote{Eigenvalues of the operator
$H(\mathcal{P}_k,\mathcal{Q}_k)$ can be parameterized by natural
numbers including 0. The $n-th$ eigenvalue is equal to $n
+\frac{1}{2}$ in units of $\hslash_E$. The lowest eigenvalue state
is called the ground state; the others are called exited states.}
state of the harmonic oscillator \cite{12}$$
W_n(p,q)dpdq=\frac{(-1)^n}{\pi\hslash_E}\thinspace
e^{-\frac{2H(p,q)}{\hslash_E\omega}}
L_n\bigl(\frac{4H(p,q)}{\hslash_E\omega}\bigr)dpdq\,,
$$ where $L_{n}$ is the $n$-th Laguerre polynomial. We can
calculate (cf\mbox{.} the definition $(\ref{defiont})$) marginal
distribution corresponding to a fixed risk strategy (that is the
associated  risk is not a random variable). We call such a
strategy an adiabatic strategy \cite{4}. The identity \cite{13}
\begin{equation*} \int_{-\infty}^\infty
\text{e}^{\text{i}kw-\frac{k^2}{4}}\,L_n
\Bigl(\frac{k^2}{2}\Bigr)\,dk=\frac{2^{n+1}\sqrt{\pi}}{n!}\,\text{e}^{-w^2}H^2_n(w)\,
,
\end{equation*} where $H_n(w)$ are the Hermite polynomials, Eq\mbox{.} $(\ref{hraddel})$ and the generating function for the Laguerre
polynomials, $\frac{1}{1-t}\thinspace
\text{e}^\frac{xt}{t-1}=\sum_{n=0}^\infty L_n(x)\,t^n$ lead to
\begin{equation}
W_{n,u,v}(w)=\frac{2^n}{\sqrt{\pi(u^2+v^2)}\,n!}\,\,\text{e}^{-\frac{w^2}{u^2+v^2}}
\,H^2_n\Bigl(\frac{w}{\sqrt{u^2+w^2}}\Bigr)=|\langle
w|\psi_n\rangle|^2 \label{hradwkw}.
\end{equation}
This is the squared absolute value of the probability amplitude
expressed in terms of the variable $w$. It should be possible to
interpret Eq\mbox{.} $(\ref{hradwkw})$ in terms of stochastic interest
rates but this outside the scope of the present paper.
\section{Canonical transformations} Let us call those
linear transformations
$(\mathcal{P},\mathcal{Q})\negthinspace\rightarrow\negthinspace(\mathcal{P'},\mathcal{Q'})$
of operators $\mathcal{P}$ and $\mathcal{Q}$ that do not change
their commutators
$\mathcal{P}\mathcal{Q}\negthinspace-\negthinspace\mathcal{Q}\hspace{.1em}\mathcal{P}$
canonical. The canonical transformations that preserve additivity
of the supply and demand components of the risk inclination
operator $\tfrac{\mathcal{P}^2}{2m}+\tfrac{m\mathcal{Q}^2}{2}\ $
\cite{4,8} can be expressed in the compact form
\begin{equation}
\begin{pmatrix}
\mathcal{P}\\\mathcal{Q}
\end{pmatrix}\begin{pmatrix}
\tfrac{\text{Re}\,z}{z\overline{z}}&\text{Im}\,z\vspace{.5ex}\\
-\tfrac{\text{Im}\,z}{z\overline{z}}&\text{Re}\,z
\end{pmatrix}
\begin{pmatrix}
\mathcal{P'}\\\mathcal{Q'}
\end{pmatrix}\,,
\label{hradcytcyt}
\end{equation}
where $z\negthinspace\in\negthinspace\overline{\mathbb{C}}$ is a
complex parameter that is related to the risk asymmetry parameter
$m$\/,  $m\negthinspace=\negthinspace z\overline{z}$. Changes in
the absolute value of the parameter $z$ correspond to different
proportions of distribution of the risk between buying and selling
transactions. Changes in the phase of the parameter $z$ may result
in mixing of supply and demand aspects of transactions. For
example, the phase shift $\tfrac{\pi}{4}$ leads to the new
canonical variables
$\mathcal{P'}=\mathcal{Y}:=\tfrac{1}{\sqrt{2}}\,(\mathcal{P}
\negthinspace-\negthinspace\mathcal{Q})$ and
$\mathcal{Q'}=\mathcal{Z}:=\tfrac{1}{\sqrt{2}}\,(\mathcal{P}\negthinspace
+\negthinspace\mathcal{Q})$. The new  variable $\mathcal{Y}$
describes arithmetic mean deviation  of the logarithm of price
from its expectation value in trading in the asset $\mathfrak{G}$.
Accordingly, the new variable $\mathcal{Z}$ describes the profit
made in one buying-selling cycle in trading in the asset
$\mathfrak{G}$. Note that the normalization if forced by the
requirement of canonicality of transformations. In the following
we will use the Schr\"{o}dinger-like picture for description of
strategies. Therefore strategies will be functions  of the
variable $y$ being the properly normalized value of the logarithm
of the market price of the asset in question. The dual description
in terms of the profit variable $z$ is also possible and does not
require any modification due to the symmetrical form of the risk
inclination operator $H(\mathcal{Y},\mathcal{Z})$ \cite{4,8}. The
player's strategy represents his/her actual position on the
market. To insist on a distinction, we will define tactics as the
way the player  decides to change his/her strategy according to
the acquired information, experience and so on. Therefore, in our
approach,  strategies are represented by  vectors in Hilbert space
and tactics are linear transformations acting on strategies (not
necessary unitary because some information can drastically change
the players behaviour!)

\section{Diffusion of prices}
 Classical description of the time
evolution of a logarithm of price of an asset is known as the
Bachelier model. This model is based on the supposition that the
probability density of the logarithm of price fulfills a diffusion
equation with an arbitrage forbidding drift. Therefore we will
suppose that the (quantum) expectation value of the arithmetic
mean of the logarithm of price of an asset $E(\mathcal{Y})$ is a
random variable described by the Bachelier model. So the price
variable $y$ has the properties of a particle performing random
walk that can be described as Brown particle at large time scales
$t$ and as Rayleigh particle at short time scales $\gamma$
\cite{16}. The superposition of these two motions gives correct
description of the  behaviour of the random variable $y$. It seems
that the parameters $t$ and $\gamma $ should be treated as
independent variables because the first one parameterizes
evolution of the "market equilibrium state" and the second one
parameterizes the "quantum" process of reaching the market
equilibrium state \cite{7,17}. Earlier \cite{14}, we have
introduced canonical portfolios as equivalence classes of
portfolios having assets with equal proportions. An external
observer describes the moves performed by the portfolio manager as
a draw in the following  lottery. Let $p_{n}, n=1,...,N$ be the
probability of the purchase of  $w_{n}$ units of the $n$-th asset.
Our analysis lead us  to Gibbs-like probability distribution:
\begin{equation}p_{n}\left( c_{0},\dots ,c_{N}\right) = \frac{  \exp \left(
\beta c_{n}w_{n} \right) } {\sum _{k=0}^{N}\exp \left(  \beta
c_{k}w_{k} \right) }\label{pkanon}.
\end{equation} The coefficient $c_{n}$ denotes the present relative price of
a unit of the asset $\mathfrak{G}_{n}$,
$c_{n}=\frac{u_{n}}{\overline{u}_{n}}$ where $u_{n}$ is the
present price of the $n$-th asset and $\overline{u}_{n}$ its price
at the moment of drawing. Now let us consider an analogue of
canonical Gibbs distribution function
\begin{equation}
\text{e}^{-\gamma H(\mathcal{Y},\mathcal{Z})}, \label{htakryk}
\end{equation}
where we have denoted the Lagrange multiplier by $\gamma$ instead
of the more customary $\beta$ for later convenience. The analysis
performed in Ref. \cite{1,15} allows to interpret
$(\ref{htakryk})$ as non-unitary tactics leading to a new
strategy\footnote{If the numbers $c_{n}w_{n}$ are eigenvalues of
some bounded below Hermitian operator $H$ then we get the
statistical operator
 $\frac{\text{e}^{ -\beta H}}{Tr \text{e}^{ -\beta H} }$. The expectation value of
 any observable $\mathcal{X}$ is given
 by $\langle\mathcal{X}\rangle_H :=\frac{Tr
 \mathcal{X}\text{e}^{ -\beta H}}{Tr \text{e}^{ -\beta H }}$.}:
\begin{equation}
\text{e}^{-\gamma H(\mathcal{Y},\mathcal{Z})}|\psi\rangle = |\psi
' \rangle. \label{dzialanie}
\end{equation}
 Therefore the parameter $\gamma$ can
be interpreted as the inverse of the temperature ($\beta \sim
(temperature)^{-1}$) of a canonical portfolio  that represents
strategies of traders having the same risk inclination (cf\mbox{.}
Ref.\cite{14}). These traders adapt such tactics  that the
resulting strategy form a ground state of the risk inclination
operator $ H(\mathcal{Y},\mathcal{Z})$ (that is they aim at the
minimal eigenvalue). We call tactics characterized by constant
inclination to risk, $E(H(\mathcal{P},\mathcal{Q}))={const}$ and
maximal entropy thermal tactics. Regardless of the possible
interpretations, adoption of the tactics $(\ref{htakryk})$ means
that traders have in view minimization of the risk (within the
available information on the market). It is convenient to adopt
such a normalization (we are free to fix the Lagrange multiplier)
of the operator of the tactics so that the resulting strategy is
its fixed point. This normalization preserves the additivity
property, $
\mathcal{R}_{\gamma_1+\gamma_2}\negthinspace=\negthinspace
\mathcal{R}_{\gamma_2}\mathcal{R}_{\gamma_1} $ and allows
consecutive (iterative) implementing of the tactics. The operator
representing such  thermal tactics takes the form
($\omega\negthinspace=\negthinspace\hslash_E\negthinspace=\negthinspace1$)
\begin{equation*}
\mathcal{R}_\gamma:=\text{e}^{-\gamma
(H(\mathcal{Y},\mathcal{Z})-\frac{1}{2})}\, .
\end{equation*}
Note that the operator $H(\mathcal{Y},\mathcal{Z})-\frac{1}{2}$
annihilate the minimal risk strategy (remember that the minimal
eigenvalue is $\frac{1}{2}$). The integral representation of the
operator $\mathcal{R}_\gamma$ (heat kernel) acting on strategies
$\langle y|\psi\rangle\negthinspace\in\negthinspace\mathcal{L}^2$
reads:

\begin{equation}
\langle y|\mathcal{R}_\gamma\psi\rangle=\int_{-\infty}^{\infty}
\negthinspace\negthinspace \mathcal{R}_\gamma(y,y') \langle
y'|\psi\rangle dy', \label{forhradof}
\end{equation}
where (the Mehler formula \cite{18})
\begin{equation*}
\mathcal{R}_\gamma(y,y')=\tfrac{1}{\sqrt{\pi(1-\text{e}^{-2\gamma})}}\,\,\text{e}^{-\frac{y^2-
{y'}^2}{2^{\vphantom{2}}}-\frac{(\text{e}^{-\gamma}y-y')^2}{1-\text{e}^{-2\gamma}}}\,.
\end{equation*}
$\mathcal{R}_\gamma(y,y')$ gives the probability density of
Rayleigh particle changing its velocity from $y'$ to $y$ during
the time $\gamma$. Therefore the fixed point condition for the
minimal risk strategy takes the form

\begin{equation*}
\int_{-\infty}^{\infty}\mathcal{R}_\gamma(y,y')\,
\text{e}^{\frac{y^2-{y'}^2}{2}}dy'=1\,.
\end{equation*}
>From the mathematical point of view, the tactics
$\mathcal{R}_\gamma$ is simply an Ornstein-Uhlenbeck process. It
is possible to construct such a representation of the Hilbert
space $\mathcal{L}^2$ so that the fixed point of the thermal
tactics corresponds to a constant function. This is convenient
because the "functional" properties are "shifted"  to the
probability measure $\widetilde{dy}\negthinspace:=\negthinspace
\tfrac{1}{\sqrt{\pi}}\,\text{e}^{-y^2}\negthinspace dy$. After the
transformation
$\mathcal{L}^2(dy)\negthinspace\rightarrow\negthinspace
\mathcal{L}^2(\widetilde{dy})$,  proper vectors of the risk
inclination operator are given by Hermite polynomials (the
transformation in question reduces to the multiplication of
vectors in $\mathcal{L}^2$ by the function
$\sqrt[4]{\pi}\,\text{e}^{\tfrac{y^2}{2}}$). Now
Eq\mbox{.} $(\ref{forhradof})$ takes the form:
\begin{equation*}
\widetilde{\langle y|\mathcal{R}_\gamma\psi\rangle}\int_{-\infty}^{\infty}\negthinspace\negthinspace
\widetilde{\mathcal{R}}_\gamma(y,y')\, \widetilde{\langle
y'|\psi\rangle}\, \widetilde{dy'}\,,
\end{equation*}
where
\begin{equation*}
\widetilde{\mathcal{R}}_\gamma(y,y'):\tfrac{1}{\sqrt{1-\text{e}^{-2\gamma}}}\,\text{e}^{{y'^2}-
\frac{(\text{e}^{-\gamma}y-y')^2}{1-\text{e}^{-2\gamma}}}\,.
\end{equation*}
In this way we get the usual description of the Ornstein-Uhlenbeck
process in terms of the kernel
$\widetilde{\mathcal{R}}_\gamma(y,y')$ being a solution to the
Fokker-Planck equation \cite{19}.
\section{"Classical" picture of quantum diffusion}
Let us consider the integral kernel of one-dimensional exponent of
the Laplace operator $\text{e}^{-\frac{\gamma}{2}\,
\frac{\partial^2}{\partial y^2}}\negthinspace$ representing the
fundamental solution of the diffusion equation
$$
\frac{\partial f(y,\gamma)}{\partial \gamma}=\frac{1}{2}\,
\frac{\partial^2 f(y,\gamma)}{\partial y^2}\,\,.
$$
The kernel takes the following form
$$
\mathcal{R}^0_\gamma(y,y'):=\tfrac{1}{\sqrt{2\pi\gamma}}\,
\text{e}^{-\frac{(y-y')^2}{2\gamma}}\,,
$$
and the appropriate measure invariant with respect to
$\mathcal{R}^0_\gamma(y,y')$  reads:
$$
dy_0:=\tfrac{1}{\sqrt{\pi\gamma}}\,
\text{e}^{-\frac{y^2}{2\gamma}}dy\,.
$$
The corresponding stochastic process is known as the
Wiener-Bachelier process. In physical applications the variables
$y$ and $\gamma$ are interpreted as position and time,
respectively (Brownian motion). Let us define the operators
$\mathcal{X}_k$ acting on $\mathcal{L}^2$ as multiplications by
functions $x_k(y(\gamma_k))$ for successive steps
$k\negthinspace=\negthinspace1,\ldots,n$ such that
$-\tfrac{\gamma}{2}\negthinspace\leq\negthinspace\gamma_1\negthinspace\leq
\negthinspace\ldots\negthinspace\leq\negthinspace\gamma_n\negthinspace
\leq\negthinspace\frac{\gamma}{2}$. The corresponding
(conditional) Wiener measure $dW^\gamma_{y,y'}$ for $
y\negthinspace=\negthinspace y(-\tfrac{\gamma}{2})$ and
$y'\negthinspace=\negthinspace y(\tfrac{\gamma}{2})$ is given by
the operator
\begin{equation*}
\int\negthinspace\prod\limits_{k=1}^n
x_k(y(\gamma_k))\,dW^\gamma_{y,y'}:=\Bigl(
\text{e}^{-\tfrac{\gamma_1+\gamma/2}{2}\tfrac{\partial^2}{\partial
y^2}} \mathcal{X}_1
\text{e}^{-\tfrac{\gamma_2-\gamma_1}{2}\tfrac{\partial^2}{\partial
y^2}} \mathcal{X}_2\cdots\mathcal{X}_n
\text{e}^{-\tfrac{\gamma/2-\gamma_n}{2}\tfrac{\partial^2}{\partial
y^2}} \Bigr)(y,y')\,.
\end{equation*}
If the operators $\mathcal{X}_k$ are constant
($x_k(y(\gamma_k))\negthinspace\equiv\negthinspace1$) then
$$
\int dW^\gamma_{y,y'}=\mathcal{R}^0_\gamma(y,y')\,.
$$
The Wiener measure allows to rewrite the integral kernel of the
thermal tactics in the form \cite{18}
\begin{equation}
\mathcal{R}_\gamma(y,y')=\int\mathcal{T}^{\prime
}\,\text{e}^{-\int \limits_{-\gamma/2}^{\gamma/2}
\frac{y^2(\gamma')-1}{2}\,\,d\gamma'}dW^\gamma_{y,y'}
\label{feynmankac}
\end{equation}
known as the Feynman-Kac formula where $\mathcal{T}^{\prime }$ is
the anti-time ordering operator. According to the quantum
interpretation of path integrals \cite{20} we can expand the
exponent function in Eq\mbox{.} $\ref{feynmankac}$ to get
"quantum" perturbative corrections to the Bachelier model that
result interference\footnote{Roughly speaking path intragrals sum
up all possible ways of evolution ("paths") with phases (weights)
resulting from interaction. } of all possible classical scenarios
of profit changes in time spread $\gamma$, cf\mbox{.}
\cite{21}.\footnote{Note that in the probability theory one
measures risk associated with a random variable by squared
standard deviation. According to this we could define the complex
profit operator
$\mathcal{A}:=\tfrac{1}{\sqrt{2}}\,(\mathcal{Y}+\text{i}\,\mathcal{Z})$.
The appropriate risk operator would take the form
$H(\mathcal{A}^\dag,\mathcal{A}) =\mathcal{A}^\dag\mathcal{A}
+\tfrac{1}{2}$.} These quantum corrections are unimportant for
short time intervals $\gamma\negthinspace\ll\negthinspace 1$ and
the Ornstein-Uhlenbeck process resembles the Wiener-Bachelier one.
This happens, for example, for "high temperature" thermal tactics
and for disorientated markets (traders)\footnote{ That is the
parameter  $\gamma$ is very small (but positive).}. In effect, due
to the cumulativity of dispersion during averaging for normal
distribution $\eta(x,\sigma^2)$
\begin{equation*}
\int_{-\infty}^\infty\negthinspace\negthinspace\eta(x\negthinspace+\negthinspace
y,\sigma^2_1)\, \eta(y,\sigma^2_2)\,dy=\eta(x,
\sigma^2_1\negthinspace+\sigma^2_2)
\end{equation*}
the whole quantum random walk parameterized by $\gamma$ can be
incorporated additively into the mobility parameter of the
classical Bachelier model. This explains changes in mobility of
the logarithm of prices in the Bachelier model that follow, for
example, from changes in the tactics temperature or received
information. Therefore the  intriguing phenomenon of market prices
evolution can be interpreted in a reductionistic way as a quantum
process. In this case the Bachelier model is a consequence of a
short-time tactics adopted by the smithonian invisible hand (under
the perfect concurrence assumption all other traders can be
considered as an abstract trader dealing with any single real
trader) \cite{4}. From the quantum point of view the Bachelier
behaviour follows from short-time interference of tactics adopted
(paths followed) by the rest of the world considered as a single
trader. Collected information about the market results after time
$\gamma\negthinspace\ll\negthinspace1$ in the change of tactics
that should lead the trader the strategy being a ground state of
the risk inclination operator (localized in the vicinity of
corrected expectation value of the price of the asset in
question). This should be done in such a way that the actual price
of the asset  is equal to the expected  price corrected by the
risk-free rate of return (arbitrage free martingale)\cite{22}.
Both interpretations of the chaotic movement of market prices
imply that Ornstein-Uhlenbeck corrections to the Bachelier model
should qualitatively matter only for large $\gamma$ scales.
 An attentive
reader have certainly noticed that we have supposed that the drift
of the logarithm of the price of an asset must be a martingale
(that is typical of financial mathematics \cite{22}). Now suppose
that we live in some imaginary state where the ruler is in a
position to decree the exchange rate between the local currency
$\mathfrak{G}$ and some other currency $\mathfrak{G'}$. The value
of the logarithm of the price of $\mathfrak{G}$ (denoted by
$\mathfrak{n}$) is proportional to the result of measurement of
position of a one dimensional Brown particle. Any owner of
$\mathfrak{G}$ will praise the ruler for such policy and prefer
$\mathfrak{G}$ to $\mathfrak{G'}$ because the the price of
$\mathfrak{G}$ in units of $\mathfrak{G'}$ will, on average, raise
(the process $\exp \mathfrak{n}$ is sub-martingale). For the same
reasons a foreigner will be content with preferring
$\mathfrak{G'}$ to $\mathfrak{G}$. This currency preference
paradoxical property of price drifts suggest that the common
assumption about logarithms of assets prices being a martingale
should be carefully analyzed prior to investment. If one measures
future profits from possessing $\mathfrak{G}$ with the anticipated
change in quotation of $\mathfrak{n}$ then the paradox is solved
and expectation values of the profits from possessing
$\mathfrak{G}$ or $\mathfrak{G'}$ are equal to zero. Therefore the
common reservations on using of logarithms of exchange rates as
martingales to avoid the Siegel's paradox is fulfilled \cite{22}
(cf\mbox{.} Bernoulli's solution to the Petersburg paradox
\cite{23}). Note that if we suppose that the price of an asset and
not its logarithms is a martingale then the proposed model of
quantum price  diffusion remains valid if we suppose that the
observer's reference system drifts with a suitably adjusted
constant velocity (in logarithm of price variable).\\
\section{Final remarks}
We have proposed a model of price movements that is inspired by
quantum mechanical evolution of physical particles. The main
novelty is to use complex amplitudes whose squared modules
describe the probabilities. Therefore such phenomena as
interference of tactics (strategies) are possible. The analysis
shows the movement of market prices imply that the Bachelier
behaviour follows from short-time interference of tactics adopted
by the rest of the world considered as a single trader and the
Ornstein-Uhlenbeck corrections to the Bachelier model should
qualitatively matter only for large time scales. Roughly speaking,
traders dealing in the asset $\mathfrak{G}$ act as a sort of
(quantum) tomograph and their strategies can be reproduced from
the corresponding Wigner functions in a way analogous to the
mathematical tomography used in medicine. Therefore we can
speculate about possibilities using  the experience acquired in
medicine, geophysics and radioastronomy to investigate intricacies
of supply and demand curves.\\
{\bf Acknowledgments} This paper
has been supported by the {\bf Polish Ministry of Scientific
Research and Information Technology} under the (solicited) grant
No {\bf PBZ-MIN-008/P03/2003}.
\def\urla{\href{http://econwpa.wustl.edu:8089/eps/get/papers/9904/9904004.html}{http://econwpa.wustl.edu:8089/eps/get/papers/9904/9904004.html}}
\def\urlb{\href{http://www.spbo.unibo.it/gopher/DSEC/370.pdf}{http://www.spbo.unibo.it/gopher/DSEC/370.pdf}}
\def\urlc{\href{http://www.comdig.org}{http://www.comdig.org}}

\end{document}